\documentclass{mn2e}
\usepackage{amsmath}
\usepackage{amssymb}
\usepackage{graphicx,epsfig}

\def\gsim{\;\lower4pt\hbox{${\buildrel\displaystyle >\over\sim}$}\;}
\def\lsim{\;\lower4pt\hbox{${\buildrel\displaystyle <\over\sim}$}\;}
\def\grls{\;\lower4pt\hbox{${\buildrel\displaystyle >\over <}$}\;}
\def\beq{\begin{equation}}
\def\eeq{\end{equation}}
\def\mpp{m_{\rm p}}
\def\hb{h_{\rm b}}
\def\np{n_{\rm p}}

\def\xx{{\rm x}}
\def\nx{n_{\rm x}}

\def\rb{\rho_{\rm b}}
\def\dd{{\rm d}}

\title[Dark Matter$-$Baryon Interaction Limits]
{Collisional interaction limits between dark\\
matters and baryons in `cooling flow' clusters}
\author[J. Hu \& Y.-Q. Lou]
{Jian Hu$^{1,2}$\thanks{E-mail:
jhu@mpa-garching.mpg.de; louyq@tsinghua.edu.cn
},
Yu-Qing Lou$^{1,3,4,5\star}$\\
$^1$Physics Department and the Tsinghua Centre for
Astrophysics (THCA), Tsinghua University, Beijing 100084, China\\
$^2$Max-Planck-Institut f\"ur Astrophysik,
Karl-Schwarzschild-Stra{\ss}e 1, 85741 Garching bei M\"unchen,
Germany\\
$^3$Centre de Physique des Particules de Marseille (CPPM)
/Centre National de la Recherche Scientifique (CNRS)\\
\qquad\quad /Institut National de Physique Nucl\'eaire et de
Physique des Particules (IN2P3) et Universit\'e\\ \qquad\ \ de la
M\'editerran\'ee Aix-Marseille II,
163, Avenue de Luminy, Case 902, F-13288 Marseille, Cedex 09, France\\
$^4$Department of Astronomy and Astrophysics, The University
of Chicago, 5640 S. Ellis Ave, Chicago, IL 60637 USA\\
$^5$National Astronomical Observatories, Chinese Academy of
Sciences, A20, Datun Road, Beijing 100012, China}
\pagerange{000,000}

\begin{document}
\date{Accepted 2007 November 22. Received 2007 October 19; in original
form 2007 April 23} \maketitle

\begin{abstract}
Presuming weak collisional interactions to exchange the kinetic
energy between dark matter and baryonic matter in a galaxy
cluster, we re-examine the effectiveness of this process in
several `cooling flow' galaxy clusters using available X-ray
observations and infer an upper limit on the heavy dark matter
particle (DMP)$-$proton cross section $\sigma_{\rm xp}$. With a
relative collisional velocity $V-$dependent power-law form of
$\sigma_{\rm xp}=\sigma_0(V/10^3\ {\rm km\ s^{-1}})^a$ where
$a\leq 0$, our inferred upper limit is $\sigma_0/m_{\rm x}\lsim
2\times10^{-25}\ {\rm cm}^2\ {\rm GeV}^{-1}$ with $m_{\rm x}$
being the DMP mass. Based on a simple stability analysis of the
thermal energy balance equation, we argue that the mechanism of
DMP$-$baryon collisional interactions is unlikely to be a stable
nongravitational heating source of intracluster medium (ICM) in
inner core regions of `cooling flow' galaxy clusters.
\end{abstract}


\section{Introduction}

Astrophysical and cosmological measurements together with numerical
simulation experiments indicate that the cold dark matter (CDM)
constitutes most of the matter in the Universe, even though the
fundamental physical nature of such dark matter particles (DMPs)
remains unknown. In the simplest scenario, these cold DMPs are
presumed to be collisionless and they interact with each other or
with other baryons only through the mutual gravity at the present
epoch. While the collisionless CDM model is successful in explaining
the formation of large-scale structures in the Universe,
observational contradictions with numerical simulations appear
inevitable regarding structures on sub-cluster scales, e.g., the
prediction of a higher number of dwarf galaxies than that actually
observed. Spergel \& Steinhardt (2000) revitalized the concept of
strongly interacting massive particles (SIMPs) to confront this
issue and suggested a self-interacting cross section per unit DMP
mass $\sigma_{\rm xx}/m_{\rm x}\sim 10^{-24}\ {\rm cm}^2\ {\rm
GeV}^{-1}$ where $\sigma_{\rm xx}$ is the collisional cross section
among DMPs and $m_{\rm x}$ is the DMP mass. Along this line of
reasoning, if DMPs are strongly self-interacting, then similar
strong interactions would also be equally expected to exist between
DMPs and baryons. Elastic scatterings between DMPs with nuclei could
generate a recoil energy of the order of $\sim 10$ keV, which would
then be detectable in underground and underwater particle
experiments. Such `direct detection' experiments currently limit the
cross section to the order of $10^{-42}-10^{-40}$ cm$^2$ for a cold
DMP of mass in the range of $\sim 10-10^3$ GeV (e.g., Akerib et al.
2004). Limits on such collisional interactions in a wider mass range
were investigated by various physical experiments and observations
of several astrophysical processes, such as $\beta\beta$ decays,
cosmic-ray detections, the galactic-halo stability, the cooling of
molecular clouds, proton decay experiments, the existence of old
neutron stars and the Earth (e.g., Starkman et al. 1990), satellite
experiments in space (e.g., Wandelt et al. 2001), primordial
nucleosynthesis and cosmic rays (e.g., Cyburt et al. 2002), and
cosmic microwave background anisotropy and large-scale structure
power spectrum (e.g., Chen et al. 2002). These experiments and
observations have provided a limit on the mass-dependent cross
section per unit DMP mass between DMPs and protons as $\sigma_{\rm
xp}/m_{\rm x}\lesssim10^{-26}-10^{-24}$ cm$^2$ GeV$^{-1}$ for a DMP
mass $m_{\rm x}\gtrsim 1$ GeV.

Recent high-resolution observations of the hot X-ray emitting
gaseous intracluster medium (ICM) by {\it Chandra} (e.g., Peterson
et al. 2001) and {\it XMM-Newton} (e.g., Kaastra et al. 2001; Tamura
et al. 2001) satellites have revealed deficits of cool gases (with
gas temperatures much less than the virial temperature $T_{\rm
vir}$) in the core of so-called `cooling flow' galaxy clusters,
inconsistent with predictions of the conventional radiative cooling
models (e.g., Cowie \& Binney 1977; Fabian \& Nulsen 1977; Mathews
\& Bregman 1978; Stewart et al. 1984; Fabian 1994 and extensive
references therein). Several heating mechanisms have been proposed
to resolve this `cooling flow dilemma', such as the inward thermal
conduction from hot outer regions (e.g., Narayan \& Medvedev 2001
and references therein), energy injections associated with central
activities by an active galactic nucleus (e.g., Churazov et al.
2002), and outward or inward acoustic wave heating (e.g., Pringle
1989; Fabian et al. 2003a, b; Fujita et al. 2004; Feng, Zhang, Lou
\& Li 2004). The resonant excitation of internal gravity modes
($g-$modes) in the ICM by orbiting galaxies was explored by Balbus
\& Soker (1990); they examined the processes of excitation,
propagation, amplification, damping of such galaxy cluster $g-$modes
in the context of providing thermal energy in `cooling flow' galaxy
clusters.
%comments on g-modes in galaxy clusters by Balbus \& Soker?
%There were analytic and simulation work of A. Malagoli et al. (1990) and
%L. Tao (1993?; MNRAS) on magnetic inhibition of thermal conduction etc.

Qin \& Wu (2001) proposed collisional interactions between baryons
and heavy DMPs ($m_{\rm x}\gg m_{\rm p}$ with $m_{\rm p}$ being the
proton mass) as a major nongravitational heating mechanism for ICM
in the core of a galaxy cluster. Assuming DMPs and baryons have
comparable velocity dispersions, the kinetic energy of a single DMP
would then be much larger than that of a baryon because of the
presumed mass difference $m_{\rm x}\gg m_{\rm p}$. Therefore in
numerous elastic collisions, kinetic energies of DMPs can be
systematically transferred to the baryonic ICM to balance the
radiative cooling in X-ray bands by hot electrons. By equating
heating and cooling rates, they estimated a specific cross section
of $\sigma_{\rm xp}/m_{\rm x}\sim 10^{-25}$ cm$^2$ GeV$^{-1}$ for
$m_{\rm x}/m_{\rm p}>10^5$. This is actually a requirement of
compensating the radiation cooling and should not be regarded as a
limit of any sort. It should also be noted that they took strong
lensing cluster CL0024+1654 to infer the DMP-baryon interaction
cross section for balancing the cooling, which may not be proper for
the following reason. The significant discrepancy by a factor of
$3\sim 4$ between the mass profiles derived from X-ray observations
and gravitational lensing effects shows that the ICM may not be in a
static equilibrium and might be still collapsing (e.g., Kneib et al.
2003; Zhang et al. 2005), and the bimodal velocity distribution of
cluster galaxies indicates a merger of two systems with a mass ratio
of $1$ to $2$ (e.g., Czoske et al. 2002). The main heating mechanism
of CL0024+1654 could well be the gravitational collapse.

Chuzhoy \& Nusser (2006) re-considered the ICM heating scenario
of Qin \& Wu (2001), corrected their calculations and derived a
similar cross section for the heavy DMP-proton elastic collisional
interaction. They found that, if $\sigma_{\rm xp}$ is independent
of the relative velocity $V$ of colliding particles, a thermal
equilibrium state between heating by DMPs and radiative cooling
by hot electrons of ICM would be always unstable. However, in
galaxy clusters with $T>2$ keV, a stable energetic balance may
be achieved for a relative velocity $V-$dependent cross section
$\sigma_{\rm xp}\propto V^a$ with $a\lesssim -3$.

There are two major simplifications in both analyses of Qin \& Wu
(2001) and Chuzhoy \& Nusser (2006). First, they took the
temperature and density of the ICM and the density of DMPs for
typical values, rather than the relevant distributions determined by
X-ray observations with high angular and spectral resolutions.
Secondly, they estimated the velocity dispersion of DMPs either
simply similar to that of baryons or by the results of numerical
simulations, rather than a dynamically self-consistent `true' value
obtained by solving the Jeans equation (Binney \& Tremaine 1987;
Subramanian 2000; see also Ikebe et al. 2004 for the case of galaxy
cluster A1795). Since the specific cross section limit they derived
(at the centre or at the virial radius) is highly sensitive to the
chosen parameters [see equation (8) of Qin \& Wu (2001)], it is
crucial to investigate this ICM heating mechanism more carefully
using an actual sample of `cooling flow' galaxy clusters, well
observed in X-ray bands by the {\it XMM-Newton} and {\it Chandra}
satellites in space.

This paper is structured as follows. In \S 2, we set the upper
limits for DMP-proton elastic collisional cross section per unit
DMP mass. In \S 3, we demonstrate that the DMP-proton collisional
interaction alone is unlikely to be a stable ICM heating mechanism
to compensate the radiative cooling of ICM. Discussion and
conclusions are contained in \S 4. Details on the X-ray
cooling function can be found in Appendix A.

In our theoretical model consideration, we have adopted the
currently favoured standard $\Lambda$CDM cosmology with the
cosmological parameters $h=0.7$, $\Omega_{\rm m}=0.3$ and
$\Omega_\Lambda=0.7$ in conventional notations.

\section{Dark Matter-Baryon Collisional\\ \ \quad
Interaction Cross Sections}

\subsection{Model Description}

%{\bf Double meanings of notations $h$, and $\Lambda$!}

The X-ray radiative cooling rate of the ICM can be inferred
observationally and may be approximately represented by the
cooling function $\Lambda$ (i.e., radiative energy loss rate per
unit volume) in terms of temperature $T$, baryon mass density
$\rho_{\rm b}$ and abundances $Z$ of the ICM as described in
Appendix A. The radiative `cooling time' $t_{\rm c}$ of a galaxy
cluster is then defined by (e.g., Sarazin 1988)
\beq\label{eq1}
t_{\rm c}\equiv\bigg|\Big(\frac{\dd\ln
\varepsilon}{\dd t}\Big)^{-1}\bigg|=
\frac{3\rho_{\rm b}k_{\hbox{\tiny B}}T/
(2\mu m_{\rm p})}{\Lambda}\ ,
\eeq
where $\varepsilon$ is the ICM internal energy per unit volume,
$k_{\hbox{\tiny B}}$ is the Boltzmann constant, $\mu$ is the mean
molecular weight of the ICM, and $m_{\rm p}$ is the proton mass.
The cooling time $t_c$ is a function of radius $r$ in general. For
`cooling flow' clusters, the estimated central cooling time is
shorter than their cosmic age, or the Hubble time $t_{\hbox{\tiny
H}}=H_0^{-1}$ with $H_0$ being the Hubble constant. The so-called
`cooling radius' $r_{\rm c}$ is defined as the radius such that
$t_{\rm c}(r_{\rm c})=t_{\hbox{\tiny H}}$.

The thermal conduction in ICM across magnetic field may be
negligible (e.g., Cowie \& Binney 1977; Stewart et al. 1984; Sarazin
1988; Fabian 1994) due to small gyro-radii given by $r_g=\gamma
mv_{\perp}c/(Z_ceB)$ where $\gamma\equiv
[1-(v_{\parallel}^2+v_{\perp}^2)/c^2]^{-1/2}$ is the relativistic
Lorentz factor, $m$ is the particle mass, $v_{\parallel}$ and
$v_{\perp}$ are the particle velocity components parallel and
perpendicular to the local magnetic field $\vec B$, $Z_ce$ is the
particle electric charge, and $c$ is the speed of light.\footnote{By
numerical simulation analysis, Tao (1993) argued however that
`tangled' magnetic field may not be effective enough to suppress the
thermal conduction in haloes of galaxy clusters.} For thermal
electrons and an average $|\vec B|\sim 1\mu G$ in a typical ICM, we
estimate $r_g$ to be a few thousand kilometers. The central regions
of `cooling flow' clusters will cool down substantially within a
timescale of $t_0$ since their formation; in the absence of other
effective heating mechanisms to compensate the radiative loss in
X-ray bands, the hot ICM core would then collapse under
self-gravity. Although the ICM behaviour in `cooling flow' clusters
is inhomogeneous on smaller scales due to thermal instabilities
(e.g., Field 1965; Mathews \& Bregman 1978; Malagoli et al. 1990),
their main large-scale properties may be grossly modelled by smooth
subsonic flows in a theoretical description (e.g., Fabian et al.
1984)\footnote{Chuzhoy \& Nusser (2006) adopted different equations
to describe the ICM behaviour. In particular, they ignored ICM flows
as well as the effect of significant gravitational heating. In these
two aspects, our model appears more general and realistic.}.
Assuming a quasi-spherical symmetry and a quasi-steady state, the
ICM evolution may be described by the following two equations,
namely, the mass conservation of baryons
\beq\label{eq2} \dot{M}=4\pi\rho_{\rm b}vr^2\
%\cong\frac{4\pi r^2\rho_{\rm b}}
%{\dd t_{\rm c}/\dd r}\bigg|_{t_{\rm c}(r)=t}\ ,
\eeq
%
%where the inflow speed $v$ of `cooling flow' is
%estimated by $dr/dt_c$ using equation (\ref{eq1}),
%{\bf precise meaning in reference to equation (\ref{eq1})? }
%\beq\label{eq3}
%\frac{\dd p}{\dd r}
%=-\rho_{\rm b}\frac{\dd\phi}{\dd r}\ ,
%\eeq
and the energy conservation of baryons in ICM (see Appendix B)
%\footnote{
%The derivation of equation 3 is described in appendix B.}
%
%\beq\label{eq4}
%\rho_{\rm b}v\frac{\dd}{\dd r}\bigg(h_{\rm b}
%+\phi+\frac{v^2}{2}
%+\frac{\langle B_t^2\rangle}{4\pi\rho_{\rm b}}\bigg)=\Lambda-H\ ,
%\eeq
%
\beq\label{eq4} \rho_{\rm b}v\frac{\partial}{\partial
r}\bigg(h_{\rm b} +\phi+\frac{v^2}{2}\bigg)=\Lambda-H\ , \eeq
where $\dot{M}$ is the baryon mass accretion rate,
%$v$ is the magnitude of the `cooling inflow' speed,
%{\bf How about the kinetic energy density?
%Subsonic `cooling flow' speed $v$? sign error?},
the ICM thermal pressure $p=\rho_{\rm b}k_{\hbox{\tiny
B}}T/(\mu\mpp)$ follows the ideal gas law, $\phi$ is the total
gravitational potential (including that of the dark matter halo),
$v^2/2$ is the baryon kinetic energy per unit mass,
%$\langle B_t^2\rangle$ is the mean square of completely
%random magnetic field $\vec B_t$ transverse to the radial
%direction (Yu \& Lou 2005; Yu, Lou, Bian \& Wu 2006; Wang \& Lou
%2007; Lou \& Wang 2007),
and $H$ represents the possible heating function (i.e., the heating
rate per unit baryon volume). The specific baryon enthalpy $h_{\rm
b}$ in energy conservation equation (\ref{eq4}) is given by
\beq
h_{\rm b}\equiv\frac{\gamma p}{(\gamma-1)\rho_{\rm b}}
=\frac{5k_{\hbox{\tiny B}}T}{2\mu\mpp}\ ,
\eeq
where the ratio of specific heats $\gamma$ is taken to be
$\gamma=5/3$ for the ICM. The estimated typical `cooling flow'
speed $v\sim 10\hbox{ km s}^{-1}$ is much less than the ICM sound
speed and thus $v^2/2$ is ignorable as compared to $h_{\rm b}$ in
energy equation (\ref{eq4}).

We could have added a magnetic term $\langle
B_t^2\rangle/(4\pi\rho_{\rm b})$ within the parentheses on the
left-hand side of energy equation (\ref{eq4}) with $\langle
B_t^2\rangle$ being the mean square of a completely random
magnetic field $\vec B_t$ transverse to the radial direction (Yu
\& Lou 2005; Yu
%, Lou, Bian \& Wu
et al. 2006; Wang \& Lou 2007; Lou \& Wang 2007). For a diffuse
ICM random magnetic field $\langle B_t^2\rangle^{1/2}$ of strength
$\sim 10^{-6}$G away from the core region (e.g., Clarke et al.
2001; Carilli \& Taylor 2002) and a typical proton number density
of $n_p\sim 10^{-3}\hbox{ cm}^{-3}$ (e.g., Sarazin 1988; Voit
2005), we estimate a typical Alfv\'en speed of $\sim 100\hbox{ km
s}^{-1}$ much less than the ICM sound speed. Such a magnetic field
strength may give rise to an anisotropic distribution of electrons
but may not be significant in the sense of bulk flow dynamics. By
the equipartition argument, a random magnetic field $\langle
B_t^2\rangle^{1/2}$ may reach strengths as strong as $\sim 10-30\
\mu$G in the ICM core region of some galaxy clusters (e.g., Fabian
1994; Hu \& Lou 2004). Depending on the actual proton number
density in the range of $n_p\sim 10^{-2}-10^{-3}\hbox{ cm}^{-3}$,
magnetic pressure is less than (e.g., Dolag \& Schindler 2000) or
may be comparable to the thermal pressure in the core region for
relaxed clusters of galaxies. For our main purpose of inferring
the upper limit of DMP mass $m_{\rm x}$ based on data of galaxy
clusters, it suffices to examine the purely hydrodynamic case.
%as done in this paper.

Now we consider the possible ICM heating mechanism due to
collisions between DMPs and baryons. As usual, we assume
Maxwellian velocity distributions for both protons and DMPs with
(one dimensional) velocity dispersions $v_{\rm p}$ and $v_{\rm x}$
respectively. For the case of $V$ independent cross section, we
have recalculated\footnote{We have verified that the constant
coefficient in equation (4) of Qin \& Wu (2001) was in error.} the
energy transfer rate per unit volume through elastic collisions as
\beq\label{transfer1} H=8\bigg[\frac{2}{\pi}\bigg(1+
\frac{v_{\rm p}^2} {v_{\rm x}^2}\bigg)\bigg]^{1/2}
\mpp\np\nx v_{\rm x}^3\sigma_{\rm xp}
\frac{1-m_{\rm p}v_{\rm p}^2/( m_{\rm x}v_{\rm x}^2)}
{(1+m_{\rm p}/m_{\rm x})^2}\ ,
\eeq
where $\np$ and $\nx$ are the number densities of protons and
DMPs, respectively. For DMPs much heavier than protons (i.e.,
$m_{\rm x}\gg m_{\rm p}$) and as $v_{\rm x}\sim v_{\rm p}$
in a gravitationally virialized system, the last factor in
the form of a division on the right-hand side of equation
(\ref{transfer1}) approaches unity.

For the case of $V$ dependent cross section $\sigma_{\rm xp}$, we
take the case that the cross section has the form of a power law,
namely $\sigma_{\rm xp}=\sigma_0(V/V_0)^a$. To avoid complicated
calculations and as an example of illustration, we only consider
the case of heavy DMPs. Adopting expression (9) of Chuzhoy \&
Nusser (2006), we would then have
%the heating rate per unit volume as
%
\beq\label{transfer2}
H\approx 6\times 2^a m_{\rm p}n_{\rm p}n_{\rm x}
v_{\rm x}^{3+a}V_0^{-a}\sigma_0(1+v_{\rm p}^2/v_{\rm x}^2)^{(1+a)/2}
\eeq
as an approximate heating function.
%{\bf The description above is not sufficiently clear.}

In our model anaylsis, the scattering cross section $\sigma_{\rm
xp}$ between a DMP and a proton is a free parameter to be
constrained by several available X-ray observations of `cooling
flow' galaxy clusters. The scattering cross section between DMPs
and helium nuclei $\sigma_{\rm xHe}$ is expected to be
$4\sigma_{\rm xp}$ for incoherent scatterings, or $16\sigma_{\rm
xp}$ for coherent scatterings, or 0 for the spin-dependent case
(e.g., Chen et al. 2002). For simplicity, we shall take
$\sigma_{\rm xHe}=0$, noting that the other two alternatives would
change the results only slightly.

The mass density and temperature distributions of the ICM can be
inferred from X-ray observations independently. The
one-dimensional velocity dispersion of ICM protons is $v_{\rm
p}=(k_{\hbox{\tiny B}}T/m_{\rm p})^{1/2}$. Under the
approximations of quasi-spherical symmetry, quasi-hydrostatic
equilibrium and ideal gas law, the total enclosed cluster mass
distribution obeys the following condition
\beq
\frac{GM_r}{r^2}=-\frac{k_{\hbox{\tiny B}}T}{\mu m_{\rm p}}\bigg(\frac{\dd\ln T}{\dd r}
+\frac{\dd\ln\rho_{\rm b}}{\dd r}\bigg)\ ,
\eeq
where $M_r$ is the total enclosed cluster mass (dark matter and
baryonic matter together) inside radius $r$, $G=6.67\times 10^{-8}
\hbox{ g}^{-1}\hbox{ cm}^{3}\hbox{ s}^{-2}$ is the gravitational
constant. The mass density of DMPs can then be inferred from
\beq
\rho_{\rm x}(r)=\frac{1}{4\pi r^2}
\frac{\dd M_r}{\dd r} -\rho_{\rm b}(r)\ .
\eeq

In most model calculations for clusters of galaxies, the DMP mass
density distribution is fitted with the universal NFW mass profile
obtained by numerical simulations (e.g., Navarro, Frenk \& White
1995, 1996, 1997), namely \beq\label{NFWdensity} \rho_{\rm
x}(r)=\rho_{\rm x0}(r/r_{\rm s})^{-1}(1+r/r_{\rm s})^{-2}\ , \eeq
where $r_{\rm s}$ is a radial scale and $\rho_{\rm x0}$ is a DMP
mass density scale when $r\cong 0.48\ r_{\rm s}$. For $r\ll r_s$,
DMP mass density $\rho_{\rm x}(r)$ scales as $r^{-1}$, while for
$r\gg r_s$, $\rho_{\rm x}(r)$ scales as $r^{-3}$. There are also
other possible dark matter mass density profiles $\rho_{\rm
x}(r)\propto r^{-l}(r+r_s)^{l-q}$ with $1\lsim l\lsim 1.5$ and
$2.5\lsim q\lsim 3$ for different combinations of inner and outer
radial scalings of DMP mass density (e.g., Moore et al. 1998, 1999;
Jing 2000; Rasia et al. 2004; Voit 2005).

In a spherical hydrostatic equilibrium with no mean streaming motions
such that $\bar v_r=\bar v_{\theta}=\bar v_{\phi}=0$, the velocity
dispersion of DMPs in a galaxy cluster and the mass distribution
is related by the Jeans equation (e.g., Binney \& Tremaine 1987;
Subramanian 2000)
\beq\label{jean} \frac{\dd}{\dd r}(\rho_{\rm
x}v_r^2) +\frac{2\beta_a}{r}\rho_{\rm x}v_r^2 +\rho_{\rm
x}\frac{GM_r}{r^2}=0\ .
\eeq
Here, the velocity anisotropy
parameter $\beta_a$ is defined as
\beq \beta_a=1-v_t^2(r)/[2v_r^2(r)]\ ,
\eeq
where $v_t^2\equiv v_\theta^2+v_\phi^2$ and $v_{r}$,
$v_{\theta}$, $v_{\phi}$ are the radial and angular velocity
dispersions with respect to the mean velocity which is almost zero
in a quasi-hydrostatic equilibrium. The formal solution of Jeans
equation (\ref{jean}) with a constant $\beta_a$ is
\beq
v_r^2(r)=\frac{r^{-2\beta_a}} {\rho_{\rm x}(r)}\int_r^\infty
x^{2\beta_a-2} \rho_{\rm x}(x)GM_x\dd x\ .
\eeq
%
%{\bf not a roman {\rm x} subscript for $M$; I have done that.}
Numerical simulations indicate a variation of the anisotropy
parameter $\beta_a$ between $\beta_a=0$ at the halo centre and
$\beta_a=0.6$ at the virial radius (e.g., Col\'{\i}n et al. 2000).
For $\beta_a$ as a function of $r$, Jeans equation (\ref{jean}) can
be integrated numerically with additional approximations (e.g.,
Subramanian 2000). As the form of $\beta_a$ cannot be determined
directly by observations, we simply set $\beta_a=0$ for an isotropic
velocity dispersion in our model analysis of galaxy clusters. In
fact, our calculations indicate that a nonzero $\beta_a$ only
affects $v_r^2$
% $v_{\rm x}^2$ {\bf you mean $v_r^2$ here?}
at a level of $\lsim 10\%$.

\subsection{The upper limit of elastic collisional
cross section constrained by X-ray observations }

Observations have shown that the ICM temperature is comparable to
the equivalent temperature of cluster galaxies (e.g., Jones \&
Forman 1984), implying that the ICM has not changed its
temperature very much from the initial hydrostatic equilibrium
state since the epoch of cluster formation. We thus consider a
more general problem of ICM heating and cooling [see equations
(\ref{eq2}) and (\ref{eq4})] in a quasi-hydrostatic manner.

Integrating equation (\ref{eq4}) from the centre $r=0$ to the
cooling radius $r_{\rm c}$ and using equation (\ref{eq2}), we
derive
\beq\label{energy}
%\bigg(h+\phi+\frac{v^2}{2} +\frac{\langle
%B_t^2\rangle}{4\pi\rho_{\rm b}}\bigg)\bigg|^{r_{\rm c}}_0
%=\int_0^{r_{\rm c}}\frac{\Lambda-H} {\dot{M}/(4\pi r^2)}\dd r\ .
\bigg(h+\phi+\frac{v^2}{2}\bigg)\bigg|^{r_{\rm c}}_0
=\int_0^{r_{\rm c}}\frac{\Lambda-H} {\dot{M}/(4\pi r^2)}\dd r\ .
\eeq
For `cooling flow' galaxy clusters, we have $\phi_0<\phi(r_{\rm
c})$ and $T_0\lesssim T(r_{\rm c})$ according to satellite X-ray
observations. Near the cluster centre, $v\sim 10\hbox{ km s}^{-1}$
is very much lower than the sound speed there. If we take either a
quasi-hydrostatic or a quasi-magnetostatic solution (Lou \& Wang
2006, 2007) for cluster evolution just before a complete
virialization or equilibrium, then $v\sim 0$ at $r=0$ and thus
$0<v^2(r_{\rm c})/2$.
%For a completely random magnetic field in
%magnetohydrodynamic (MHD) evolution (e.g., Zel'dovich \& Novikov
%1971), the frozen-in condition leads to $\langle
%B_t^2\rangle\propto\rho_{\rm b}^2r^2$ (Yu \& Lou 2005; Yu et al.
%2006; Lou \& Wang 2007) and thus $\langle
%B_t^2\rangle/(4\pi\rho_{\rm b})\propto\rho_{\rm b}r^2$; for an
%almost flat central $\rho_{\rm b}$ or a sufficiently slow fall-off
%of central $\rho_{\rm b}$ (e.g., $\rho_{\rm b}\propto r^{-1}$), the
%Alfv\'en speed squared $\langle B_t^2\rangle/(4\pi\rho_{\rm b})$ at
%$r=r_{\rm c}$ is larger than that at $r=0$.
By the above consideration and estimates of three terms at both
$r=r_{\rm c}$ and $r=0$ separately, we show the left-hand side
(LHS) of energy equation (\ref{energy}) is positive.\footnote{Note
that in reference to a large radius, $h+\phi+v^2/2$ should be
negative for a gravitationally bound system, while the difference
$(h+\phi+v^2/2)_{r_c}-(h+\phi+v^2/2)_{0}$ can be positive as
shown.} In fact, this requirement may be relaxed a bit. In fact,
as long as the LHS of energy equation (\ref{energy}) is
non-negative and irrespective of the relative magnitudes of each
individual terms at both $r=r_{\rm c}$ and $r=0$, we have for the
baryonic ICM the following inequality for the net heating rate
\beq\label{ine14}
\int(\Lambda-H)r^2\dd r> 0\ ,
\eeq
and the collisional cross section is limited from above by
\beq\label{ine15}
\sigma_{\rm xp}<\int_0^{r_{\rm c}}\Lambda r^2\dd
r\bigg/\int_0^{r_{\rm c}}H_1r^2\dd r\ ,
\eeq
where $H_1\equiv H/\sigma_{\rm xp}$ according to expression
(\ref{transfer1}), or by
\beq\label{ine16} \sigma_0 V_0^{-a}<\int_0^{r_{\rm c}}\Lambda r^2\dd
r\bigg/\int_0^{r_{\rm c}}H_2r^2\dd r\ , \eeq
where $H_2\equiv HV_0^{a}/\sigma_0$ according to expression
(\ref{transfer2}). By the above analysis and reasoning for the
roles of flow kinetic energy and magnetic energy and in reference
to inequalities (\ref{ine14})$-$(\ref{ine16}), our model
calculations and X-ray data comparisons can be much simplified
within a purely quasi-hydrostatic framework and that is what we do
in the following.
\begin{table*}
\begin{center}
\begin{minipage}{140mm}
\caption{Estimated physical parameters for a sample
of five `cooling flow' clusters of galaxies\label{ccs}}
\begin{tabular}{lccccccc}
\hline Cluster & $z$ & $T$ (keV) & $c_{200}$ & $R_{200}$ (kpc) &
$M_{200}$ $(10^{14}{\rm M}_\odot)$ & $r_{\rm c}$(kpc) & $\dot{M}$
(M$_\odot$ yr$^{-1}$)\footnote{The dark matter mass deposition rates
from `cooling flows'
%column 7 {\bf
(column 8) are taken from White et al. (1997). While their model
calculations used cosmological parameters of $h=0.5$ and $q_0=0.5$,
the corresponding parameter values here are all comparable to
theirs.
%{\bf the relative value is comparable.?}
}\\
\hline
 A478 & 0.0881 & 6.8 & $4.22\pm0.39$ & $2060\pm110$
& $10.8\pm1.8$ & $171^{+113}_{-115}$ & $736^{+114}_{-434}$ \\
 A1795 & 0.0616 & 5.7 & $4.47\pm0.27$ & $1760\pm30$
& $6.54\pm0.35$ & $129^{+80}_{-87}$ & $321^{+168}_{-213}$ \\
 A1983 & 0.0442 & 2.3 & $3.83\pm0.71$ & $1100\pm140$
& $1.59\pm0.61$ & $34^{+59}_{-34}$ & $6.0^{+10.8}_{-6.0}$ \\
 A1991 & 0.0586 & 2.6 & $5.78\pm0.35$ & $1106\pm41$
& $1.63\pm0.18$ & $52^{+67}_{-12}$ & $37^{+36}_{-11}$ \\
 PKS0745 & 0.1028 & 7.0 & $7.05\pm0.28$ & $1880\pm130$
& $8.34\pm0.84$ & $126^{+22}_{-31}$ & $579^{+399}_{-215}$\\
\hline
\end{tabular}
\end{minipage}
\end{center}
\end{table*}

We calculate the DMP-proton elastic collisional cross section
limit using X-ray observations of a sample of five presumably
relaxed galaxy clusters with apparent `cooling' cores, namely,
Abell 478 (e.g., Pointecouteau et al. 2004), Abell 1795 (e.g.,
Ikebe et al. 2004), Abell 1983 (e.g., Pratt \& Arnaud 2003), Abell
1991 (e.g., Pratt \& Arnaud 2005), and PKS 0745-191 (e.g., Chen et
al. 2003; Pointecouteau et al. 2005). Physical parameters
estimated for this galaxy cluster sample are summarized in Table
1. This sample of galaxy clusters covers wide ranges of redshift
$z$ ($0-0.1$), cluster virial mass $M_{200}$
($[1-11]\times10^{14}$ M$_\odot$), and ICM temperature $T$ ($2-7$
keV), respectively. These chosen galaxy clusters are analysed with
the deprojection technique based on the X-ray data taken from the
{\it XMM-Newton} EPIC to reveal the real spectra of gaseous ICM in
different spherical shells and to determine the deprojected
temperature and the mass distribution of the gas in a galaxy
cluster. The {\it XMM-Newton} EPIC is the most sensitive X-ray
space telescope which also has high spatial and spectral
resolutions, and thus meets all the requirements for a detailed
spectral analysis. The statistical study on the comparison of
cluster mass measurements using strong gravitational lensing and
using X-ray observations shows an excellent mutual agreement for
`cooling flow' galaxy clusters, which are perceived as dynamically
more relaxed systems and the quasi-spherical hydrostatic
assumption used in X-ray mass determinations should be valid
(Allen 1998; Wu 2000). On this ground, the dark matter mass
density distribution used in this paper is regarded as reliable,
especially for the lensing galaxy cluster PKS 0745-191 (e.g.,
Allen 1996).

For the velocity independent case as given by expression (\ref{transfer1}),
the upper limits of the specific cross section $\sigma_{\rm xp}/m_{\rm x}$
for DMP-proton collisions are displayed in Figure 1. The limits of
$\sigma_{\rm xp}/m_{\rm x}$ for the five selected sample galaxy clusters
lie in the range of [2.0, 4.9]$\times10^{-25}$ cm$^2$ GeV$^{-1}$ and a
common consistent value would be taken as $\sigma_{\rm xp}/m_{\rm x}
\lsim 2\times10^{-25}$ cm$^2$ GeV$^{-1}$. For the velocity dependent
case as given by expression (\ref{transfer2}), the value of
$\sigma_0/m_{\rm x}$ is displayed in Figure 2. Here, we take the mean
value of cooling radius $r_c$ of every cluster for simplicity. The
upper limit of DMP specific cross section is $\sigma_{\rm xp}
/m_{\rm x}<2\times10^{-25}(V/10^3\ {\rm km\ s^{-1}})^a$ cm$^2$ GeV$^{-1}$.

Note that $\sigma_{\rm xp} /m_{\rm x}$ is constant only for $m_{\rm
x}\gg m_{\rm p}$ in expressions (\ref{transfer1}) and (\ref{transfer2}).
%{\bf Not obvious in equation (\ref{transfer2})!}
Our estimated upper
limit applies well to $m_{\rm x}>10^3$ GeV. The cross-section limit
for the lighter DMPs ($m_{\rm x}=10-10^3$ GeV) has been strictly
constrained by direct detection experiments (e.g., Wandelt et al.
2001).

\begin{figure}
\centerline{\includegraphics[scale=.7]{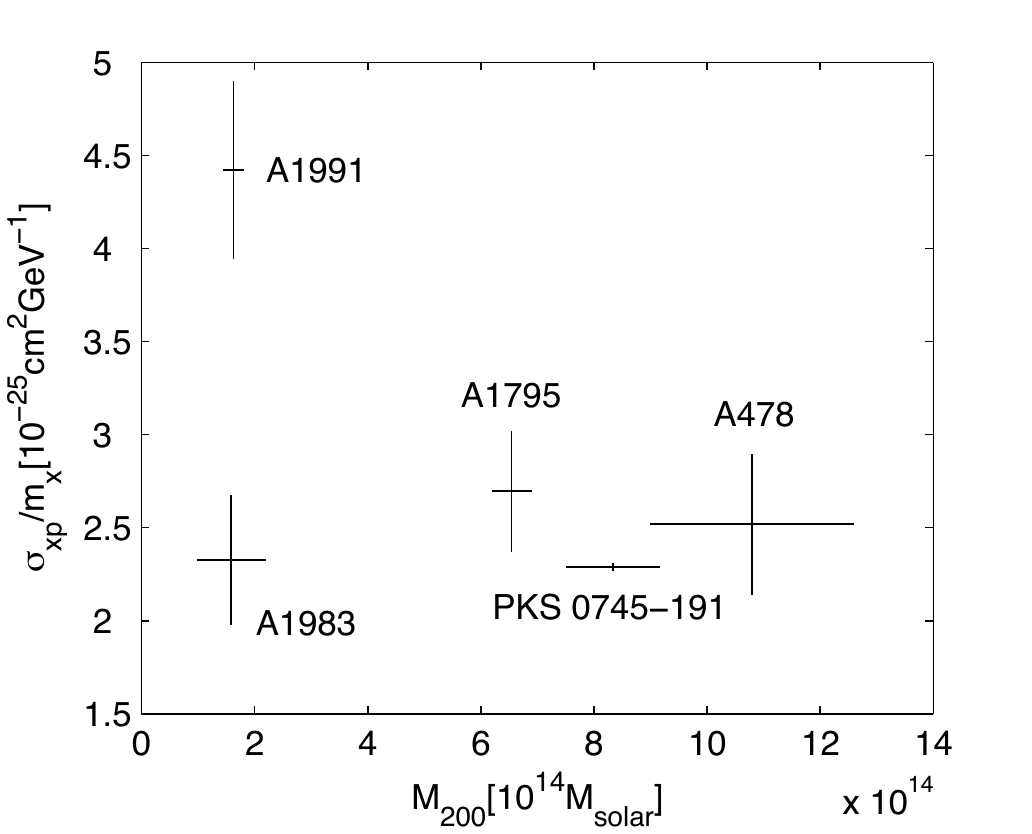}} \caption{On the
basis of data analysis and for the case of expression
(\ref{transfer1}), the upper limits of specific cross section
$\sigma_{\rm xp}/m_{\rm x}$ for the five selected `cooling flow'
galaxy clusters (namely, PKS 0745-191, Abell 478, Abell 1795,
Abell 1991, and Abell 1983) are shown here. The value range of the
estimated upper limit for $\sigma_{\rm xp}/m_{\rm x}$ is mainly
due to uncertainties of the cooling radius $r_c$ and the virial
mass.}
\end{figure}

\begin{figure}
\centerline{\includegraphics[scale=.7]{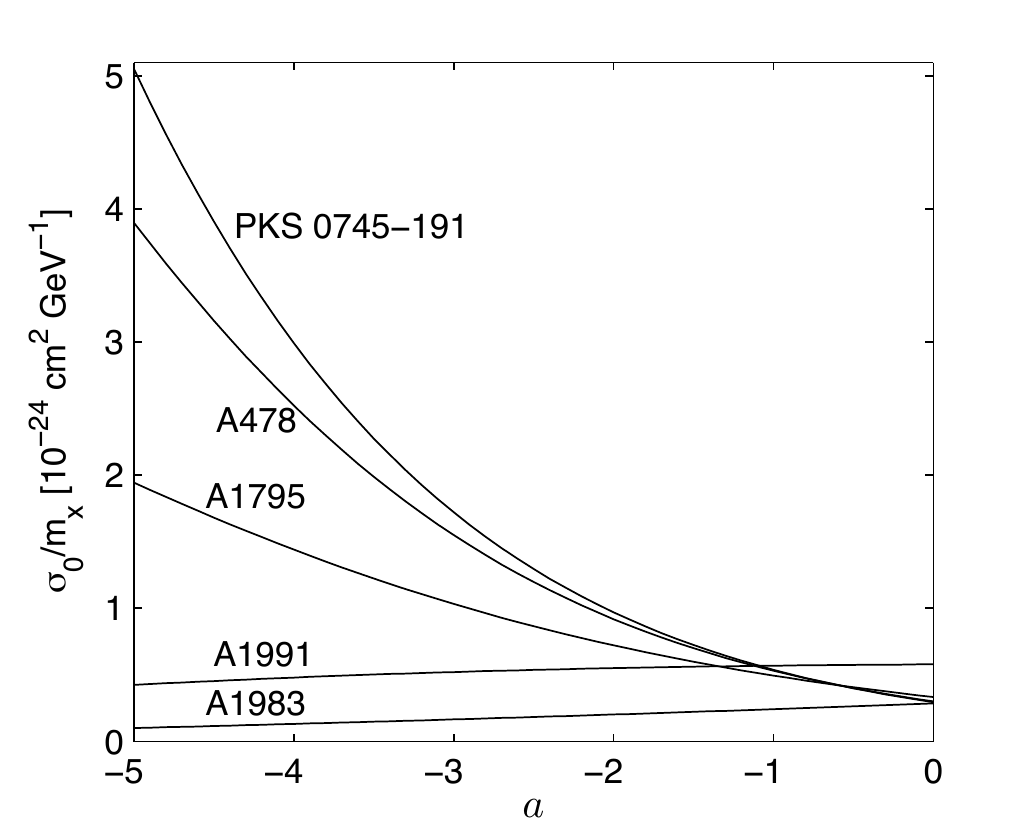}} \caption{For the
relative velocity $V-$dependent case of expression
(\ref{transfer2}), the upper limits of cross section
$\sigma_0/m_{\rm x}$ as functions of exponent $a$ for the five
selected `cooling flow' galaxy clusters (namely, PKS 0745-191,
Abell 478, Abell 1795, Abell 1991, and Abell 1983) are shown here
by five separate solid curves with corresponding galaxy cluster
names labelled explicitly.}
\end{figure}

\section{Dark Matter$-$Proton Collisions as the ICM
Heating Source in `Cooling Flow' Clusters of Galaxies}

The energy equilibrium equation of the ICM can be written as
\beq\label{br}
\dd\ln S/\dd t =\dd\ln\varepsilon/\dd t=(H-\Lambda)/\varepsilon\ ,
\eeq
where $S=k_{\hbox{\tiny B}}T/(\mu m_{\rm p}\rb^{2/3})$ is the ICM
entropy (e.g., Voit 2005).
%\footnote{This definition of entropy is
%related to but differs from the conventional definition of entropy
%in textbooks of thermodynamics.}
We note that the ratio $v_{\rm p}/v_{\rm x}$ in two expressions
(\ref{transfer1}) and (\ref{transfer2}) is almost a constant in
dynamically and thermally  well relaxed galaxy clusters (Hu \& Lou
2007). Hence for a specific `cooling flow' cluster, $H\propto\rb
T^0$ and $\Lambda\propto\rb^2T^b$, and equation (\ref{br}) becomes
\beq\label{eq18}
\dd\ln(T/\rb^{2/3})/\dd t=AT^{-1}-B\rb T^{b-1}\ ,
\eeq
where $A$ and $B$ are two coefficients independent of $\rb$ and
$T$. A systematic comparison between the ICM temperatures $T$ and
velocity dispersion of DMP base on the X-ray observations
indicates that $T$ could only slightly change ($\Delta T\lesssim$1
keV) in the cosmic age of clusters (Hu \& Lou 2007), which is
consistent with the numerical simulations of the cluster evolution
(e.g., Bryan \& Norman 1998). If the RHS of equation (\ref{eq18})
maintains zero, i.e. cooling is fully compensated by heating, we
have $\dd\log\rb=-b\dd\log T$, where $|b|\lesssim 1$ for $T\sim
10^7-10^8$ K. Therefore $\rb$ will be roughly a constant in such a
case.

Now we fix the value of $T$ and perform an isothermal stability
analysis on the solution $\rb$ of equation (\ref{eq18}). Suppose
the DMP-proton elastic collisional heating can compensate
radiative cooling for certain values $T=T_0$ and $\rb=\rho_0$, let
us consider the stability of $\rb$ in equation (\ref{br}) at a
constant temperature. A small baryon density variation $\delta\rb$
obeys
\beq \frac{2}{3\rb}\frac{\dd \delta\rb}{\dd t}=BT^{b-1}\delta\rb\
. \eeq
%
%Obviously, the time evolution of $\rb$ is unstable.
A simple estimation indicates that $\delta\rb$ will double in a
timescale of
\begin{eqnarray}
&&2^{-a}\times 10^9 {\rm yr}\ \bigg(\frac{v_\xx}{\rm 10^3
\ km\ s^{-1}}\bigg)^{-1-a}\nonumber \\
&&\ \ \ \ \times\bigg(\frac{\rho_\xx}{\rm 0.02\ M_\odot\ pc^{-3}}
\frac{\sigma_0/m_\xx}{\rm 10^{-25}\ cm^2\ GeV^{-1}}\bigg)^{-1}\ .
\end{eqnarray}
Therefore, without other heating sources, the solution of equation
(\ref{br}) appears unstable for whatever values of $a$, at least
in the case of a fixed ICM temperature profile.

\section{Discussion and Conclusions}

Based on two different elastic collision models and X-ray
observations of galaxy clusters, we have re-examined effects of
collisional interactions between heavy DMPs and protons in
`cooling flow' clusters of galaxies and estimated a more reliable
upper limit for the specific cross section between DMPs and
protons by solving the equation of energy conservation for five
``cooling flow" galaxy clusters using available X-ray data. In the
regime of $m_{\rm x}\gg m_{\rm p}$, the upper limit for
$\sigma_{\rm xp}=\sigma_0(V/10^3\ {\rm km\ s^{-1}})^a$ with
$a\leq0$ can be expressed as $\sigma_0/m_{\rm x}\lsim 2\times
10^{-25}$ cm$^2$ GeV$^{-1}$, which is fully consistent with the
earlier results. Similar to other astrophysical constraints, this
upper limit is independent of the underlying model for particle
physics governing collisional interactions.

In our model calculations, we have assumed that the total
gravitational potential $\phi$ remains invariable during
the process of galaxy cluster evolution. The gravitational
potential $\phi$ of a galaxy cluster is dominantly determined
by the amount of DMPs. Besides the weak energy-momentum transfer
between the DMPs and baryons, the density distribution of dark
matter is only affected by the mild accumulation of cool gas
in the central region of clusters. The exact solution of such
complex co-evolution of DMPs and baryons should be treated with
Fokker-Planck equation (e.g., Binney \& Tremaine 1987). We will
pursue the relevant problems in contexts of galaxy clusters
%(e.g., Lou 2005)
in separate papers.

We have also demonstrated an intrinsic instability in explaining
the problem of `cooling flow' galaxy clusters via the heat transfer
by heavy DMP-proton collisional scattering. According to expression
(\ref{transfer1}), the DMP-proton collisional heating of ICM works
only when $m_{\rm p}v_{\rm p}^2<m_{\rm x}v_{\rm x}^2$. In `cooling
flow' galaxy clusters, $v_{\rm p}^2/v_{\rm x}^2\sim\mu\simeq 0.6$
(Hu \& Lou 2007) and thus $m_{\rm x}>0.6$ GeV is required in this
model. However, as the collisional cross section in the DMP mass
range 0.5 GeV$<m_{\rm x}<10^5$ GeV is less than $5\times10^{-28}$
cm$^2$ (e.g., Wandelt et al. 2001), the light DMP collisional
heating mechanism can then be completely ignored. Therefore, we
rule out the possibility of DMP collisional scattering as a major
stable non-gravitational heating sources of the ICM in `cooling
flow' galaxy clusters.

\section*{Acknowledgments}
This research has been supported in part
by the NSFC grants 10373009 and 10533020 at the
Tsinghua University,
%by the ASCI Center for Astrophysical Thermonuclear
%Flashes at the University of Chicago,
by the Special Funds for Major State Basic Science Research
Projects of China, by THCA,
%by the Collaborative Research Fund from the National
%Science Foundation of China for Outstanding Young
%Overseas Chinese Scholars (NSFC 10028306) at the National
%Astronomical Observatories, Chinese Academy of Sciences,
and by the SRFDP 20050003088
%Specialized Research Fund for the Doctoral
%Program of Higher Education
and the Yangtze Endowment from the Ministry of Education to
the Tsinghua University. The hospitalities of the Mullard Space
Science Laboratory at University College London, U.K., of School
of Physics and Astronomy, University of St Andrews, Scotland, U.K.,
and of Centre de Physique des Particules de Marseille (CPPM/IN2P3/CNRS)
et Universit\'e de la M\'editerran\'ee Aix-Marseille II, France are
also gratefully acknowledged.
%Affiliated institutions of YQL share this contribution.

\appendix

\section{X-ray Cooling Function}

The radiative energy loss from a hot plasma per unit volume and per
unit time is a cooling function of temperature, density, and chemical
abundances of various elements. The cooling functions for a hot plasma
under equilibrium conditions have been calculated over a photon energy
range of $0.001-30$ keV and a range of abundances for $0-1.0\ Z_\odot$
by Sutherland \& Dopita (1993). They defined the normalized cooling
function by $\Lambda_0\equiv\Lambda/(n_{\rm e} n_{\rm i})$, where
$n_{\rm e}$ and $n_{\rm i}$ are the electron and ion number densities,
respectively. Tozzi \& Norman (2001) used an approximate analytic
expression to fit $\Lambda_0$ in the following polynomial form
\beq\label{lambdan}
 \Lambda_0=[C_1(k_{\hbox{\tiny B}}T)^{-1.7}
+C_2(k_{\hbox{\tiny B}}T)^{0.5}+C_3] \times10^{-22}\ {\rm erg}\ {\rm
cm}^3 \ {\rm s}^{-1}\ ,
\eeq
where $k_{\hbox{\tiny B}}T$ is in unit of keV. The three constant
coefficients $C_i$ ($i=1,2,3$) and $n_{\rm e}/n_{\hbox{\tiny H}}$,
$n_{\rm i}/n_{\hbox{\tiny H}}$, $\mu$ depend on the metallicity $Z$.
We have refitted these parameters and summarized the results in Table
\ref{tabcf} for the convenience of reference. For an arbitrary
metallicity $Z$, we take the parameters by the following polynomial
interpolations:
\begin{eqnarray}
 C_1&=&-0.003+0.26Z-0.41Z^2+0.29Z^3\ ,\\
 C_2&=&0.0605+0.034Z-0.069Z^2+0.052Z^3\ ,\\
 C_3&=&0.019+0.041Z-0.025Z^2\ ,\\
 n_{\rm e}/n_{\hbox{\tiny H}}&=&1.12+0.14Z-0.05Z^2\ ,\\
 n_{\rm i}/n_{\hbox{\tiny H}}&=&1.06+0.06Z-0.02Z^2\ ,\\
 \mu&=&0.575+0.074Z-0.030Z^2\ .
\end{eqnarray}
Expression (\ref{lambdan}) with parameters defined by
equations (A2)$-$(A7) can grossly reproduce the X-ray
cooling function of Sutherland \& Dopita (1993) to
within a few percent in the typical ICM thermal energy
range of 1.0 keV $\lsim k_{\hbox{\tiny B}}T\lsim$ 10 keV.

\begin{table}
%\centering
 \begin{minipage}{140mm}
\caption{Parameters for the Normalized
Cooling Function $\Lambda_0$
%(Appendix A)
\label{tabcf}}
 \begin{tabular}{lllllll}
 \hline
$Z$ ($Z_\odot$) & $C_1$ & $C_2$ & $C_3$ &
$n_{\rm e}/n_{\hbox{\tiny H}}$ & $n_{\rm i}/n_{\hbox{\tiny H}}$ & $\mu$\\
  \hline
 0 & $-0.003$ & $0.0605$ & $0.0204$ & 1.128 & 1.064 & 0.58 \\
 0.1 & 0.0193 & 0.0632 & 0.0218 & 1.131 & 1.064 & 0.58 \\
 $0.32$ &$0.480$ & $0.0658$ & $0.0306$ & 1.165 & 1.080 & 0.60 \\
 $1.0$ & $0.1434$ & $0.0762$ & $0.0355$ & 1.209 & 1.099 & 0.62\\
  \hline
\end{tabular}
\end{minipage}
\end{table}

\section{Derivation of energy conservation for ICM baryons}

The energy conservation equation for ICM baryons (equation 3) is
derived in this Appendix B.
We start with the thermodynamic relation
\beq\label{b1} \dd
\hb=\rb^{-1}\dd p+\dd Q=\rb^{-1}\dd p+\rb^{-1}(H-\Lambda)\dd t\ ,
\eeq
where $\hb$ is the enthalpy per unit baryon mass, $Q$ is the
heating rate per unit baryon mass, $p$ is the ICM (baryon) gas
pressure, and $\Lambda$ and $H$ are cooling and heating rates per
unit volume, respectively. Differentiating along the moving
direction of a bulk baryon flow, equation (\ref{b1}) can be written
as
\beq\label{b2} \textbf{v}\cdot(\rb^{-1}\triangledown
p)=\textbf{v}\cdot\triangledown \hb+\rb^{-1}(\Lambda-H)\ .\eeq
Using the Euler equation (e.g., Landau \& Lifshitz 1959) \beq
\frac{\partial \textbf{v}}{\partial
t}+(\textbf{v}\cdot\triangledown)\textbf{v}= -\rb^{-1}\triangledown
p-\triangledown\phi\ , \eeq where $\phi$ is the total gravitational
potential (including that of a dark matter halo), and a vector
identity \beq
(\textbf{v}\cdot\triangledown)\textbf{v}=\triangledown\bigg(\frac{v^2}{2}\bigg)
-\textbf{v}\times (\triangledown\times\textbf{v})\ ,
%=\triangledown\frac{v^2}{2}
\eeq equation (\ref{b2}) appears as \beq
\textbf{v}\cdot\frac{\partial \textbf{v}}{\partial t}=
-\textbf{v}\cdot\triangledown\bigg(\frac{v^2}{2}+\phi+\hb\bigg)
+\frac{(H-\Lambda)}{\rb}\ . \eeq Here we assume that the moving
direction of a cooling flow is radially inward (i.e., negative),
such that \beq\label{b6} \frac{\partial v}{\partial
t}=\frac{\partial}{\partial
r}\bigg(\frac{v^2}{2}+\phi+\hb\bigg)+\frac{(H-\Lambda)}{\rb v}\
.\eeq
We now estimate and compare the magnitudes of terms in equation
(\ref{b6}). For example, \beq \frac{\partial v}{\partial
t}\sim\frac{v}{t_{\rm H}}\sim\frac{10\ \rm km\ s^{-1}}{10^{10}\
\rm yr}\sim3\times10^{-14}{\rm m\ s^{-2}}\ , \eeq
\beq \frac{\partial \phi}{\partial
r}=\frac{GM_r}{r^2}\gtrsim\frac{GM_r}{r^2}\Big|_{r=R_{200}}
\sim\frac{10^{14}\ \rm M_\odot G}{(1\ \rm Mpc)^2}\sim 10^{-11}
{\rm m\ s^{-2}}\ , \eeq
\beq \frac{\partial}{\partial
r}\bigg(\frac{v^2}{2}\bigg)\sim\frac{v^2}{2r}\sim
\frac{(10\hbox{km/s})^2}{1\hbox{Mpc}}\sim 3\times 10^{-15}\hbox{m
s}^{-2}\ , \eeq
\beq \frac{\partial h_b}{\partial r}\sim \frac{h_b}{r} \sim
\frac{5\ (1 \hbox{keV})}{2\mu m_p\ 1\hbox{Mpc}}\sim 10^{-11}{\rm
m\ s^{-2}}\ ,\eeq
where $GM_r/r^2$ is a decreasing function with increasing $r$ for
NFW mass profiles obtained through numerical simulations. The
value of $(H-\Lambda)/(\rho_b v)$ term is to be determined, yet it
should be much larger than terms $\partial (v^2/2)/\partial r$ and
$\partial v/\partial t$ in equation (\ref{b6}). Therefore the LHS
and $\partial(v^2/2)/\partial r$ term on the RHS of equation
(\ref{b6}) may be neglected and we finally derive equation (3) as
\beq \rb v\frac{\partial}{\partial
r}\bigg(\frac{v^2}{2}+\phi+\hb\bigg)=\Lambda-H\ . \eeq

\end{document}